# The GAPS Experiment to Search for Dark Matter using Low-energy Antimatter

R. A. Ong[1,2], T. Aramaki[3], R. Bird[2], M. Boezio[4], S.E. Boggs[5], R. Carr[6], W.W. Craig[7], P. von Doetinchem[8], L. Fabris[9], F. Gahbauer[10], C. Gerrity[8], H. Fuke[11], C.J. Hailey[10], C. Kato[12], A. Kawachi[13], M. Kozai[11], S.I. Mognet[14], K. Munakata[12], S. Okazaki[11], G. Osteria[15], K. Perez[6], V. Re[16], F. Rogers[6], N. Saffold[10], Y. Shimizu[17], A. Yoshida[18], T. Yoshida[11], G. Zampa[4], and J. Zweerink[2]

The GAPS experiment is designed to carry out a sensitive dark matter search by measuring low-energy cosmic ray antideuterons and antiprotons. GAPS will provide a new avenue to access a wide range of dark matter models and masses that is complementary to direct detection techniques, collider experiments and other indirect detection techniques. Well-motivated theories beyond the Standard Model contain viable dark matter candidates which could lead to a detectable signal of antideuterons resulting from the annihilation or decay of dark matter particles. The dark matter contribution to the antideuteron flux is believed to be especially large at low energies ($E < 1$ GeV), where the predicted flux from conventional astrophysical sources (i.e. from secondary interactions of cosmic rays) is very low. The GAPS low-energy antiproton search will provide stringent constraints on less than 10 GeV dark matter, will provide the best limits on primordial black hole evaporation on Galactic length scales, and will explore new discovery space in cosmic ray physics.

Unlike other antimatter search experiments such as BESS and AMS that use magnetic spectrometers, GAPS detects antideuterons and antiprotons using an exotic atom technique. This technique, and its unique event topology, will give GAPS a nearly background-free detection capability that is critical in a rare-event search. GAPS is designed to carry out its science program using long-duration balloon flights in Antarctica. A prototype instrument was successfully flown from Taiki, Japan in 2012. GAPS has now been approved by NASA to proceed towards the full science instrument, with the possibility of a first long-duration balloon flight in late 2020. This presentation will motivate low-energy cosmic ray antimatter searches and it will discuss the current status of the GAPS experiment and the design of the payload.

[1]Speaker
[2]University of California, Los Angeles
[3]SLAC National Accelerator Laboratory
[4]INFN, Sezione di Trieste
[5]University of California, San Diego
[6]Massachusetts Institute of Technology
[7]Lawrence Livermore National Laboratory
[8]University of Hawaii at Manoa
[9]Oak Ridge National Laboratory
[10]Columbia University
[11]Japan Aerospace Exploration Agency
[12]Shinshu University
[13]Tokai University
[14]Pennsylvania State University
[15]INFN, Sezione di Napoli
[16]Università di Bergamo
[17]Kanagawa University
[18]Aoyama Gakuin University

## 1. Introduction

The existence of dark matter as the dominant gravitational mass in the universe is now well established from many different measurements, but its actual nature is still completely unknown. Present observations indicate that dark matter is non-baryonic and is compatible with a collisionless fluid of particles that interact only very weakly with ordinary matter. Many theories postulate the existence of a new stable, relatively heavy particle – the weakly interacting massive particle (WIMP). An attractive feature of the WIMP is that its existence can arise naturally from new physics that is invoked to explain some of the outstanding problems in the standard model of particle physics (e.g. hierarchy problem, unification of gauge couplings, etc.). Thus, for many reasons, the nature of dark matter is one of the most compelling mysteries facing physics and astronomy today.

It is now well appreciated that the identification of dark matter will likely require several complementary approaches [1], including the direct detection of dark matter via its interaction in underground detectors, new physics searches at collider experiments (e.g. at the Large Hadron Collider), and the indirect detection of dark matter using the secondary particles produced in dark matter annihilation or decay. Indirect detection experiments typically use photon (gamma-rays, X-rays), neutrino or cosmic ray signatures. Cosmic ray antiparticles (e.g. positrons, antiprotons, and antideuterons) are promising candidates for the indirect detection of dark matter, especially if the fluxes from astrophysical sources (be it primary, secondary or tertiary) are expected to be low or negligble.

In recent years, some of the most interesting hints for dark matter have come from indirect detection experiments. A gamma-ray excess observed in Fermi-LAT data from the Galactic Center region was interpreted as evidence for 30-50 GeV dark matter [2], although explanations using astrophysical sources have also been put forward [3]. The Fermi-LAT team recently reported results from the Andromeda galaxy in which a clear and extended gamma-ray excess is now seen with a spatial distribution that is similar to what is observed in the Milky Way [4]. A second recent paper by Fermi-LAT provides an updated analysis of the Galactic Center region [5]; the excess reported earlier is confirmed but enhancements of similar amplitude are found in nearby control regions indicating some sort of systematic bias or incomplete knowledge of the diffuse gamma-ray background. The latest data from AMS [6] indicate a flattening in the positron fraction, followed by a possible downturn. The increasingly well-measured excess of positrons is not understood and could originate from dark matter. The AMS antiproton energy spectrum [7] is not fully consistent with standard propagation models using updated measurements of the proton and He fluxes and the B/C ratio, and this has also been interpreted as evidence for dark matter [8]. Finally, the intriguing report of a few candidate antihelium events in the AMS data [9] has prompted speculation as to their origin [10]. All of these hints are tantalizing, but also underscore the need for experiments using complementary detection techniques and especially "smoking gun" signatures that are as background free as possible.

## 2. Low-energy Antideuterons and Antiprotons: Unexplored Phase Space

Antideuterons have never been detected in the cosmic rays, but the great promise that they could offer for the indirect detection of dark matter was well-appreciated and first discussed more than 15 year ago [11]. In WIMP annihilation or decay, standard model particles are created, including antiprotons and antineutrons, which can coalesce to form an antideuteron. Antideuterons created this way throughout the Galaxy can propagate to Earth since the amount of interstellar material they encounter is modest (~5 g/cm$^2$). Secondary or tertiary (background) antideuterons can be created when cosmic ray hadrons interact with the interstellar medium, but the background level is expected to be very low for a number of reasons. First, in a fixed-target collision of this sort, the kinematics strongly disfavor the formation of antideuterons with low kinetic energy. Second, the relatively high production threshold needed to create an antideuteron means that higher-energy projectiles are required and, given the steeply falling cosmic ray energy spectrum, this in turn leads to a substantial reduction in the secondary/tertiary flux.

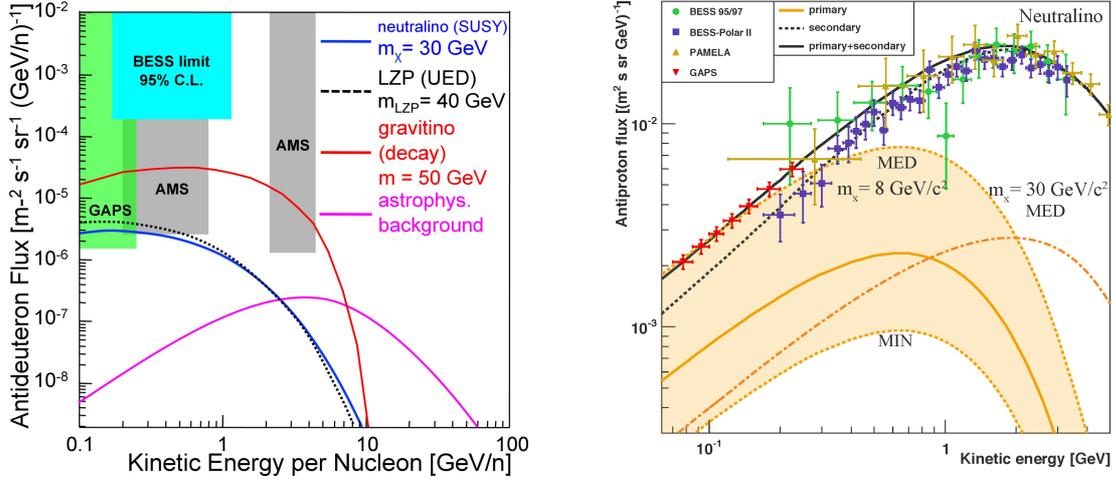

*Figure 1: Left: antideuteron flux as a function of kinetic energy per nucleon. The expected GAPS sensitivity (99% CL) [12] after three 35-day flights is shown, along with an estimation for the AMS sensitivity for 5 years of data [13] and the upper limit obtained by BESS [14]. Also shown are the predicted antideuteron fluxes for three different WIMP models and for the secondary background; see [12] for details. Right: antiproton fluxes measured by BESS and PAMELA as a function of kinetic energy and the expected flux precision for GAPS from one 35-day flight [15]. Also shown are the predicted contributions from two neutralino models under various assumptions on the propagation. See [15] for details as well as predictions from gravitino, Kaluza-Klein and primordial black hole models.*

Figure 1 (left) shows the antideuteron flux predicted from several benchmark dark matter models compared to the predicted level of the secondary/tertiary background. Also shown is the sensitivity of the GAPS experiment for three 35-day long duration balloon (LDB) flights in the Antarctic. The great potential of low-energy antideuterons can be seen, where the flux from dark matter exceeds the background level by more than two orders of magnitude. This is in marked contrast to the situation for positrons or high-energy antiprotons where the dark matter contribution sits on top of a large (and very uncertain) astrophysical background. Note also that the predictions for the signal antideuterons from dark matter in Figure 1 make conservative assumptions on any possible boost and on the propagation model. The signal level could well be higher by factors of several, increasing the possibility for a GAPS discovery.

The Alpha Magnetic Spectrometer (AMS), launched in 2011, is the only currently-operating experiment with significant antideuteron search capability. AMS probes a higher energy region than GAPS and it uses a standard magnetic spectrometer technique. GAPS is fully complementary to AMS in that it probes the lowest energies and uses a completely different exotic atom technique. In addition, the systematic uncertainties for GAPS as a balloon payload in the Antarctic are expected to be significantly different from those for AMS on the International Space Station in mid-latitude (e.g. high-geomagnetic cutoff) orbit. The importance of complementary experiments in search for rare events as a potential signal for dark matter is well recognized in the direct-detection community where more than a dozen experiments are taking data or being planned. Finally, it is important to note that independent of any particular models, GAPS and AMS will explore considerable phase space by being more than two orders of magnitude more sensitive than the best previous experiment, as shown in Figure 1.

Beyond antideuterons, GAPS has significant potential for precision measurement of low-energy antiprotons. At kinetic energies below 0.25 GeV, the statistical precision of the BESS and PAMELA data that have been used to constrain dark matter models is low. With a single LDB flight, GAPS will detect an order of magnitude more low-energy antiprotons than these previous experiments or AMS [15]. Figure 1 (right) shows the expected flux precision for GAPS, along with the existing antiproton data and the predicted contributions from two neutralino models. GAPS is particularly effective at probing light dark matter which is still viable in a number of scenarios, in spite of being severely constrained by direct-detection experiments. See [15] for more details on the GAPS antiproton capabilities as well as the predicted sensitivity to primordial black hole (PBH) evaporation and to dark matter models incorporating gravitinos and Kaluza-Klein right-handed neutrinos.

## 3. The GAPS Detection Technique and Instrument Concept

The GAPS experiment is designed to detect low-energy antideuterons and antiprotons during several LDB campaigns in Antarctica. The basic detection technique is shown in Figure 2 (left). A low-energy antiparticle crosses two layers of a time-of-flight (TOF) system composed of plastic scintillator detectors. The TOF system measures the incoming particle velocity and deposited energy (dE/dx) which can be used to estimate a stopping depth for the antiparticle in the target. The antiparticle slows down and stops in a target that is made up of layers of Si(Li) detectors. The stopped antiparticle forms an exotic atom that decays on a time scale of nanoseconds. Atomic X-rays are emitted as the atom de-excites, followed by pion and proton emission from the nuclear annihilation. The Si(Li) tracker determines the trajectories of the incoming antiparticle and the hadronic tracks emerging from the annihilation star. In addition, it measures the energies of the emitted X-rays with a resolution of several keV.

A critical capability for GAPS is to reliably detect antideuterons and antiprotons, separating both from any potential backgrounds and separating one from another. The characteristics of the exotic atom technique make very clean separations possible. In the decay of the exotic antidetueron atom, the emitted X-rays have unique energies (30 keV, 44 keV and 67 keV), as shown in Figure 2, and the average number of particles (pions and protons) emerging from the nuclear star is approximately 7.2. The toplogy of an antideuteron event, combined with these unique emissions, is exceedingly difficult to mimick by background proceses (e.g. stopped low-energy protons). Similarly, antiproton events can be easily separated from antideuteron events because they contain a different set of characteristic X-rays (23 keV, 35 keV and 58 keV) and a smaller number of particles from the nuclear star (3.0 on average). Moreover, the stopping range of antideuterons is about twice as large as that of antiprotons for the same velocity. A prototype detector, tested at the KEK accelerator in Japan in 2004-2005, confirmed the principle of the exotic atom technique [16]. The same technique can also be applied for the identification of low-energy antihelium with a larger dE/dx energy loss and a greater number of particles expected from the nuclear star. More dedicated studies are needed.

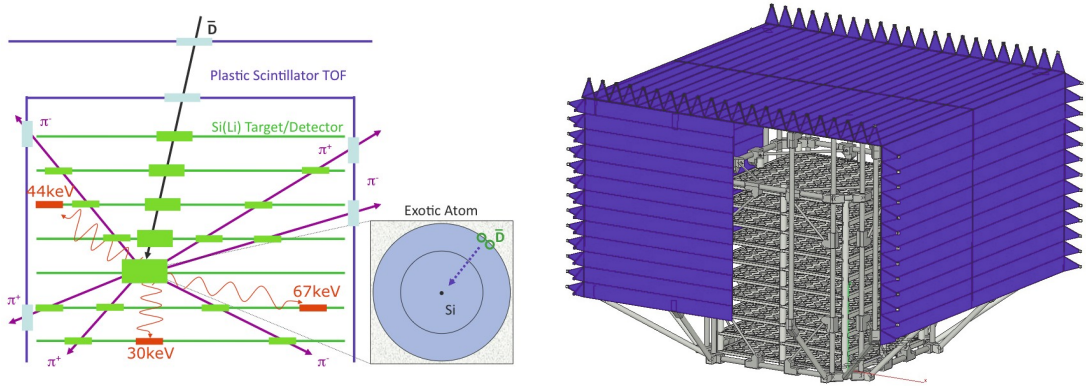

*Figure 2: Left: the GAPS detection method. An antideuteron crosses two layers of a time-of-flight (TOF) detector and slows down and stops in the Si(Li) target, forming an exotic atom. Atomic X-rays are emitted at the atom de-excites, followed by the production of pions (and protons) from the nuclear annihilation. Right: the GAPS detector design. Ten layers of Si(Li) detectors are surrounded the inner and outer TOF scintillation paddles. The inner TOF layers are not shown in order to see the target.*

The GAPS detector design is shown in Figure 2 (right). The target/tracker comprises ten layers of 4-inch diamter, 2.5mm thick Si(Li) detectors, with each detector segmented into four strips. Adjacent tracking layers will be separated by 20cm and have their strips positioned orthogonally. The inner TOF system (1.6m x 1.6m x 2.0m) will be composed of plastic scintillator detectors that completely surround the tracker. The outer TOF system (3.6m x 3.6m x 2.0m), also composed of scintillation detectors, will cover the top half of the inner detector and will be separated from the inner TOF system by 1.0m.

## 4. The GAPS Instrument Design and Current Status

The design of GAPS employs several unique features. It uses the exotic atom technique and it will be the first balloon instrument with a large Si(Li) detector and a very large TOF system without a pressure vessel. The hardware for GAPS has been in development for a number of years and a successful prototype instrument (pGAPS) was flown in 2012 from Taiki, Japan [17]. Here we discuss the ongoing development of the science instrument.

### 4.1 Si(Li) Detectors

Lithium-drifted silicon (Si(Li)) detectors are key to the success of the GAPS detection technique. The Si(Li) detectors must be 2.5mm thick in order to slow and stop the incident antiparticle, and ten planes provide sufficient depth information to distinguish antiparticles with different mass (when combined with the TOF). A modest ~4 keV energy resolution is sufficient to distinguish X-rays from exotic atoms produced by different antiparticle species. To identify the tracks resulting from pions and protons, each 4-inch diameter detector is divided into four readout strips. In contrast, modern semiconductor technology has focused on thinner detectors with increasingly precise energy resolution and fine spatial resolution, such as are required for tracking detectors in high-energy particle collider experiments.

The preliminary fabrication scheme and testing procedure have been demonstrated at a custom-built facility at Columbia University. Both 1mm- and 2mm-thick detectors, with effective areas up to 2-inch diameter, have been produced. The 2-inch detectors have demonstrated the required ~4 keV energy resolution at the temperatures foreseen in flight (-35° C), and testing using cosmic rays has verified the expected signal amplitude for minimum-ionizing particles. This work validated that detectors with sufficient performance can be produced using the lithium-drifting technique.

In 2015, we began a collaboration with Shimadzu Corp., a commercial producer of Si(Li) detectors. Shimadzu will use their facilities and staff to produce the ~1350 flight detectors. We have worked to transfer our detector technology and geometry, aiming to optimize the high-volume fabrication scheme and expand to the 4-inch diameter, 2.5mm-thick, four-strip geometry. Shimadzu has recently produced the first detector with the required geometry and with a leakage current that is close to the design requirement (see Figure 3, left). Experience has shown that the noise is very sensitive to chemical surface treatments, the condition of the Si/Li layer underneath the readout contacts, and the uniformity of the diffused Li layer after drifting and before guard ring and strip segmentation. Controlling these details and optimizing detector yield are the goals in the upcoming months.

**4.2 Si(Li) Electronics**

The readout of the Si(Li) detectors need to achieve an energy resolution of 4 keV (FWHM) over the X-ray range and provide at the same time a 60-70 MeV upper energy scale. This needs to be accomplished within a power budget (corresponding to ~110 mW for a full readout channel in the present baseline design). With pGAPS, we flew a preamplifier that consumed 15 mW and that would meet the requirements for GAPS. A power budget of ~95 mW/channel for the remaining functions of the Si(Li) data acquisition clearly pushes the limits of what can be achieved using a discrete off-the-shelf component design, and we thus opted to use integrated technologies (i.e. an ASIC) to implement the post-amplification and digitization functions. We are, however, still exploring a fully integrated circuit where the preamplifier would be part of the readout ASIC.

Once the preamplifier architecture has been chosen, the remaining functions of the readout chain are rather conventional: each preamplifier signal will be shaped to optimize signal-to-noise and its amplitude will be digitized. Each ASIC will have a serial digital data link to move the digitized information to an internal FPGA card. The card will further serialize the data to minimize the number of feedthroughs required, and it will also implement other functions such as addressing, event word building and level 0 trigger generation based on signals provided by the TOF system. The data from the FPGA card will be sent outside of the detector vessel to the on-board computer for processing, storage and telemetry.

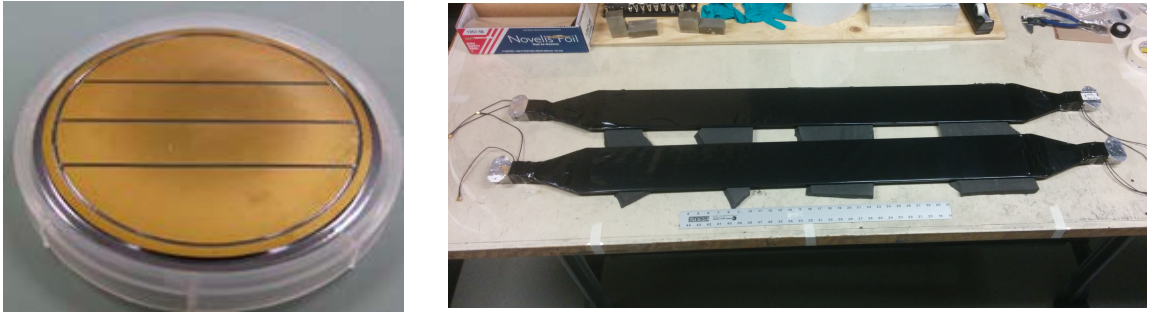

*Figure 3: Left: a prototype 4-inch diameter Si(Li) detector with the required geometry for GAPS (courtesy Shimadzu Corp). Right: A prototype 1.2m long TOF scintillation counter that uses PMTs and light guides.*

**4.3 Time-of-Flight (TOF) System**

The TOF measures the incoming particle's velocity and energy deposition, provides the high-speed trigger, and serves as a shield/veto for the instrument. The TOF is very large and thus the constraints on its weight become important. To achieve the needed mass separation, a TOF resolution of 500ps is required; this resolution is moderate compared to typical particle physics experiments and hence thinner scintillator of 5mm thickness can be used to save weight. It is important to note that GAPS will trigger on events with $0.2 < \beta < 0.5$, where the scintillation light emitted will be 3-6 times higher than for minimum-ionizing tracks. The baseline design of the TOF uses long scintillation counters that have dimensions of 180cm x 16cm in the outer TOF system and 160cm x 16cm in the inner TOF system. Approximately 220 scintillation counters are required in total.

Two approaches are being evaluated for the detection of the scintillation light. The baseline approach uses curved acrylic light guides on both ends of the scintillator to couple the light to high-speed 30mm x 30mm photomultiplier tubes (PMTs, Hamamatsu R7600-UBA). This approach was used successfully in pGAPS. We are also evaluating the use of silicon photomultipliers (Si-PMs), where three 6mm x 6mm Si-PMs would be mounted directly on each end of the scintillator. Figure 3 (right) shows a prototype scintillation counter employing PMTs and light guides. In either approach, the photosensor signal will be sampled by a high-speed (1 or 2 GS/s) digitizer ASIC that will be packaged into a custom readout board that is currently under development. The readout boards will be mounted directly on the outside of the TOF planes to minimize the cable lengths. A separate, single digital board using a high-speed FPGA is being designed to make the master trigger decision for GAPS. The trigger needs to veto a large fraction of the high rate (~100 kHz) of downward going cosmic rays passing through the apparatus and at the same time be as sensitive as possible to the signal events, which also have particle tracks from the nuclear annihilation. This is a challenging task given the relatively coarse granularity of the TOF system.

## 5. Summary

In the quest to understand dark matter, searching for low-energy antideuterons is a very promising, but largely unexplored, technique. The GAPS experiment, a large-acceptance cosmic ray balloon instrument, will push the sensitivity limit for antideuterons by two orders of magnitude compared to the best present-day measurements. The development of the science payload of GAPS is now well underway with a potential first flight occurring in late 2020.

## Acknowledgements

This work is supported in the U.S. By NASA APRA grants (NNX17AB44G, NNX17AB45G, NNX17AB46G, and NNX17AB47G), in Japan by MEXT/JSPS KAKENHI grants (JP26707015, JP17H01136, and JP17K14313), and in Italy by Istituto Nazionale di Fisica Nucleare (INFN). R.A. Ong receives support from the UCLA Division of Physical Sciences. K. Perez receives support from the Heising-Simons Foundation and RCSA Cottrell College Science Award ID #23194. P. von Doetinchem receives support from the National Science Foundation under award PHY-1551980.